# A low-energy compact Shanghai-Wuhan electron beam ion trap for extraction of highly charged ions

Shiyong Liang,[1,2,3] Qifeng Lu,[4,5] Xincheng Wang,[6] Yang Yang,[4,5] Ke Yao,[4,5] Yang Shen,[4,5] Baoren Wei,[4,5] Jun Xiao,[4,5,*] Shaolong Chen,[1,2,3] Pengpeng Zhou,[1,2,3] Wei Sun,[1,2] Yonghui Zhang,[1] Yao Huang,[1,2] Hua Guan,[1,2,†] Xin Tong,[1] Chengbin Li,[1] Yaming Zou,[4,5] Tingyun Shi,[1,7] Kelin Gao[1,2,7]

[1] State Key Laboratory of Magnetic Resonance and Atomic and Molecular Physics, Wuhan Institute of Physics and Mathematics, Chinese Academy of Sciences, Wuhan 430071, China
[2] Key Laboratory of Atomic Frequency Standards, Wuhan Institute of Physics and Mathematics, Chinese Academy of Sciences, Wuhan 430071, China
[3] University of Chinese Academy of Sciences, Beijing 100049, China
[4] Institute of Modern Physics, Department of Nuclear Science and Technology, Fudan University, Shanghai 200433, China
[5] Key Laboratory of Nuclear Physics and Ion-Beam Application (MOE), Fudan University, Shanghai 200433, China
[6] School of Physical Science and Technology, ShanghaiTech University, Shanghai 201210, China
[7] Center for Cold Atom Physics, Chinese Academy of Sciences, Wuhan 430071, China

Corresponding authors: * xiao_jun@fudan.edu.cn, † guanhua@wipm.ac.cn

## ABSTRACT

A low-energy, compact and superconducting electron beam ion trap (the Shanghai-Wuhan EBIT or SW-EBIT) for extraction of highly charged ions is presented. The magnetic field in the central drift tube of the SW-EBIT is approximately 0.21 T produced by a pair of high-temperature superconducting coils. The electron-beam energy of the SW-EBIT is in the range of 30–4000 eV, and the maximum electron-beam current is up to 9 mA. Acting as a source of highly charged ions, the ion-beam optics for extraction is integrated, including an ion extractor and an einzel lens. A Wien filter is then used to measure the charge-state distribution of the extracted ions. In this work, the tungsten ions below the charge state of 15 have been produced, extracted, and analyzed. The charge-state distributions and spectra in the range of 530–580 nm of tungsten ions have been measured simultaneously with the electron-beam energy of 279 eV and 300 eV, which preliminarily indicates that the 549.9 nm line comes from $W^{14+}$.

## I. INTRODUCTION

An electron beam ion trap (EBIT) is a device that can produce and trap highly charged ions (HCIs).[1,2] EBIT is capable of producing any possible charge-state ions of any element.[3,4] Furthermore, the tunability and quasi-monoenergeticity of its electron-beam have empowered the research of energy-dependence atomic processes of HCIs.[5] Over the past several decades, EBIT has been used to study electron-impact excitation (IE),[2] X-ray spectroscopy,[6] dielectronic recombination (DR),[7–11] radiative recombination (RR),[11] and quantum electrodynamic effects.[12–14]

In early times, super EBITs were designed and operated with the electron-beam energy ranging from a few keV to 100 keV or higher.[1,15–25] Most of the reported experiments at these times were intended to produce high-charge-state ions, i.e., $Ba^{46+}$,[2,26] $Sc^{19+}$,[6] $I^{50+}$,[7] and $U^{86-92+}$.[4,11] Afterwards, some new interests were raised on the spectroscopy of moderate-charge-state ions that could be found in both astrophysical and laboratory plasmas. For example, iron (Fe) and nickel (Ni) ions of the charge states around 10 are abundant in the corona.[27] Tungsten (W) ions of the charge states below 30 are also expected in the edge or divertor area of the next-generation fusion device ITER.[28] The spectra of such ions are significant for diagnosing these plasmas.[28–31] Hence several low-energy EBITs[32–36] were developed, and several high-energy EBITs[37–40] were operated at low-energy scales, in response to these needs.

These experiments described above were conducted on the ions trapped in EBIT. However, an EBIT integrated with ion-extraction optics can operate as an efficient source of HCIs.[41] Up to now, many mini-EBITs have been developed. The advantages of these mini-EBITs are that they are lower cost, easier to operate, and in particular, they can be used as portable HCIs sources,[42–49] for the applications such as tumor ion therapy,[5,50] ion implantation,[5] ion lithography,[5] precision spectroscopy,[51,52] and especially, the new-generation atomic clocks[53] based on HCIs proposed by Berengut et al. in 2010.[54] The energy level structure of an HCI is insensitive to external perturbations because of the size of its electron cloud being more compact due to much stronger nuclear constraint.[5] Therefore, HCIs that have appropriate clock transitions are ideal for optical atomic clocks with possible $10^{-19}$ or even smaller uncertainty level.[55,56] Besides, some HCI clock transitions are very sensitive to the fine structure constant α, which means that it is possible to test the time variation of α with smaller uncertainty.[54,56] In 2001, Gruber et al. already injected $Xe^{44+}$ ions into a cryogenic Penning trap[57] and sympathetically cooled them down to below 1.1(2) K with laser-cooled $Be^+$ ensemble for precision spectroscopy.[51,52] In 2015, Schmöger et al. retrapped single $Ar^{13+}$ ions extracted from Hyper-EBIT in a cryogenic linear Paul Trap[58] and cooled them down to below 221(26) mK by sympathetic cooling with four laser-cooled $Be^+$ ions.[59,60] Sympathetic cooling of one $Ar^{13+}$ ion by one $Be^+$ ion was also demonstrated, which is the prerequisite for quantum logic spectroscopy[61] of an optical atomic clock with a potential of reaching a $10^{-19}$ level of accuracy.[60]

For the purposes of developing ultra-precise optical atomic clocks and searching for a variation of the fine structure constant, we report on a low-energy compact EBIT, named Shanghai-Wuhan EBIT (SW-EBIT), acting as an HCIs source. The SW-EBIT is the upgraded version of the SH-HtscEBIT[35] with the capability of ion extraction. The maximum electron beam energy of the SW-EBIT is 4 keV that is capable of producing most proposed HCIs candidates for optical atomic clocks as in Refs. 62–64. The design of the SW-EBIT and the first operation results, including spectroscopy and ion extraction, are described in Secs. II and III. Some conclusions are given in Sec. IV.

## II. DESIGN

### A. Overview of the SW-EBIT

The SW-EBIT consists of an electron gun, three-section drift tubes (DT1, DT2, DT3), an electron collector, an ion extractor, a magnetic shield, an einzel lens, two high-temperature superconducting coils (magnetic coils), a liquid nitrogen ($LN_2$) tank, and a main vacuum chamber, as shown in FIG. 1. The size of the SW-EBIT is 60 cm (height) × 35 cm (outer diameter), and the main vacuum chamber and $LN_2$ tank are made of 304 stainless steel. The material of magnetic shield is soft iron coated with nickel. The oxygen-free high-conductivity copper (OFHC) electrodes and the sapphire insulators provide an excellent performance of electrical insulation and thermal conduction. The lengths of DT1, DT2, and DT3 are 4 mm, 24 mm, and 4 mm respectively, and their inner diameters are 4 mm. There are four 20 mm×1 mm slits in DT2 for gas injection and spectroscopy measurement. An approximately 0.21 T magnetic field in the center of DT2 is produced by a pair of high-temperature superconducting coils immersed in liquid nitrogen. The electron beam emitted from the cathode passes through the drift tubes and is collected by the electron collector. The electron collector is magnetically shielded in order to collect an electron beam effectively. When the gaseous sample is injected, ions are produced in DT2. These ions are radially trapped by the space-charge potential of the electron beam, and axially trapped by the electrostatic well formed by the biases of drift tubes. If necessary, the HCIs can be extracted by raising the potential of DT2 or by lowering the potential of DT3. When extracted, these HCIs are accelerated by the potential difference between DT3 and the electron collector but radially they are confined by the electron beam. They are then extracted immediately by a more negative potential on the ion extractor. Before being delivered to the next beamline from the top of the SW-EBIT, the HCI beam is focused by the einzel lens in order to optimize the beam diameter and divergence angle.

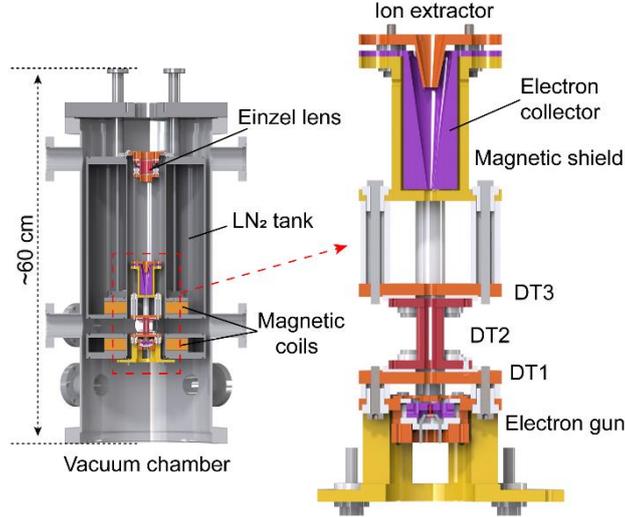

**FIG. 1**. Structure of the SW-EBIT. HCIs are produced and trapped in DT2. In the figure, LN$_2$ tank stands for liquid nitrogen tank, and DT1–3 stand for drift tube 1–3 respectively.

## B. Magnetic field design and simulation

The magnetic field of the SW-EBIT is produced by a pair of high-temperature superconducting coils immersed in liquid nitrogen. The distance between the two centers of the coils is 57.2 mm, and the size of a single coil is 62 mm (inner diameter) × 127 mm (outer diameter) × 18 mm (length). The magnetic field was simulated using the finite-element method software COMSOL.[65] The axial simulative magnetic flux density of the coil current of 20 A is shown in **FIG. 2(b)**. The magnetic flux density in the center of DT2 is approximately 0.21 T, and the magnetic field is asymmetric because of the influence of the soft iron magnetic shield. The magnetic field non-uniformity along the axis of DT2 is approximately 8%. The electron collector is shielded by a soft iron magnetic shield with the minimum magnetic flux density below 11 G.

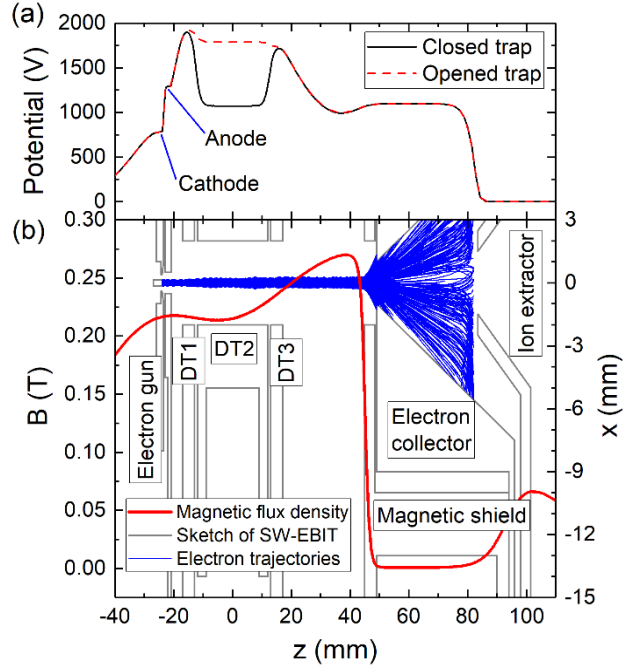

**FIG. 2**. Simulation of the electric field, magnetic field, and electron trajectories using COMSOL. (a) A typical axial electric field distribution of opened and closed traps, which contains the space-charge effect of the electron beam. (b) The axial magnetic flux density, the electron trajectories of the closed trap, and a sketch of the SW-EBIT. The current of the magnetic coils is 20 A and the electron-beam current is 4.5 mA.

## C. Simulation of electron trajectories

The cathode of the electron gun is lanthanum hexaboride (LaB$_6$) crystal (Kimball Physics ES-423-E 90-300), with its emission surface being 0.3 mm in diameter. There is no magnetic shield outside the cathode. Thus, the cathode is exposed to the strong magnetic field. With the electrostatic field shown in **FIG. 2(a)** (the closed trap) and the magnetic field shown in **FIG. 2(b)**, the electron trajectories of a 4.5 mA electron beam were simulated using COMSOL, as shown in **FIG. 2(b)**. Before arriving at the electron collector, the electron beam diameter is 0.32 mm, which includes more than 90% electrons along the path. The electron beam diverges quickly in the electron collector with the help of the magnetic shield and the repelling of the ion extractor. Finally, the electrons are collected by the electron collector. Our design mentioned above can minimize the necessary potential difference between the electron collector and the cathode, which can reduce the heating power of the electron collector.

## D. Simulation of ion trajectories

The trajectories of $^{184}W^{14+}$ ions were also simulated by COMSOL with the magnetic and electric fields shown in **FIG. 2** (the opened trap). In SW-EBIT, for the 4.5 mA electron-beam current, the maximum radial trap depth for $^{184}W^{14+}$ ions is approximately 168 eV, indicating that the real temperature should be below 2 MK. The $^{184}W^{14+}$ ions with the temperature of 0.5 MK, 1 MK, and 1.5 MK were released in a small cylinder (diameter: 0.6mm, length: 1 mm) in the center of DT2 with the Maxwell-Boltzmann velocity distribution. The ions escaped from DT3 are accelerated by the potential difference between DT3 and the electron collector, and they fly toward the electron collector with a small diameter because of the radial confinement by the space-charge potential of the electron beam. Most of the ions are extracted by the more negative potential on the ion extractor once they have arrived at the electron collector. The einzel lens is designed to focus the ion beam because the extracted ion beam diverges severely when passing a long distance. For the temperature of 1 MK, at a distance of 550 mm from the center of DT2 (the entrance of the Wien filter described in Sec. III.B), the ion beam diameter is 3.4 mm (FWHM) as shown in FIG. 3(b) and the divergence angle is 8.6 mrad (FWHM) as shown in FIG. 3(c). There is no significant difference of the simulated ion-beam diameter and divergence angle for the temperatures of 0.5 MK, 1 MK, and 1.5 MK, but the extraction ratio is smaller for higher temperatures because more ions hit on DT2 and the ion extractor, as shown in TABLE I.

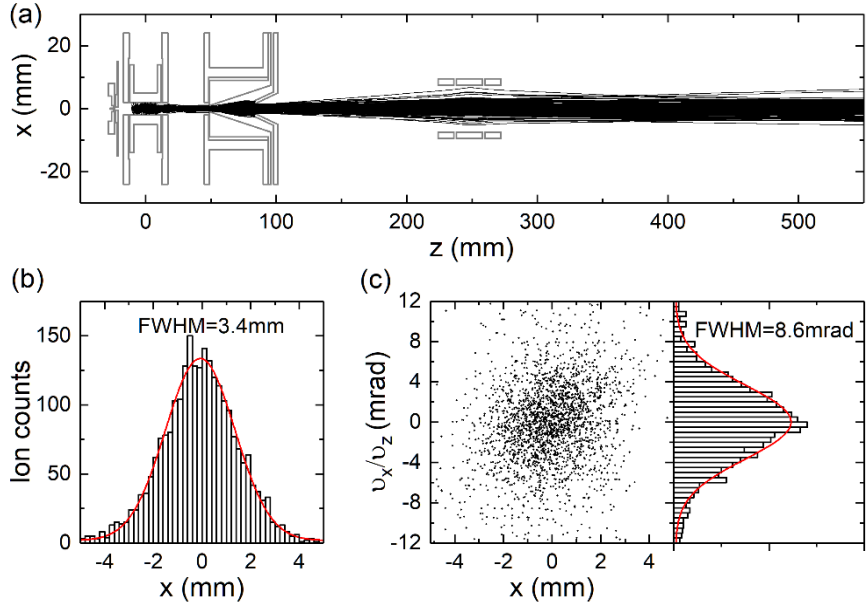

**FIG. 3**. Trajectory simulation of $^{184}W^{14+}$ ions with the temperature of 1 MK. (a) The trajectories of $^{184}W^{14+}$ ions. (b) The radial position distribution of $^{184}W^{14+}$ ions at z = 550 mm. (c) A phase space plot and ion direction ($v_x/v_z$) distribution of $^{184}W^{14+}$ at z = 550 mm.

TABLE I. The simulated ion-beam parameters of 0.5 MK, 1 MK, and 1.5 MK at z = 550 mm.

| Temperature (MK) | 0.5 | 1 | 1.5 |
| --- | --- | --- | --- |
| Diameter (mm) | 3.1 | 3.4 | 3.4 |
| Divergence angle (mrad) | 8.3 | 8.6 | 9.2 |
| Extraction ratio | 66% | 50% | 39% |

### E. Pumping and gas-injection system

The SW-EBIT's main chamber is pumped from the bottom by a 300 L/s turbo molecular pump backed by an oil-free scroll pump. A one-stage ultra-high vacuum (UHV) differential system (gas chamber) is used to inject gaseous samples into the center of DT2, which is connected to a 90 L/s turbo molecular pump backed by the same oil-free scroll pump. A UHV needle valve is used to adjust the injection rate to the gas chamber. Typically, the pressure of the gas chamber ranges from $10^{-7}$ to $10^{-6}$ Torr when the gas is injected. A 27.5 cm long tube with 4 mm inner diameter (2 mm at the nozzle) is used to separate the gas and the main chamber, achieving 3–4 orders of magnitude pressure difference. Therefore, the pressure of the main chamber does not change significantly, when the gas-injection system is working.

## III. OPERATION AND MEASUREMENTS

### A. Overview of operation

The volume of the liquid nitrogen tank is approximately five liters, and the liquid nitrogen consumption rate is approximately one liter per hour. With the help of low temperature, the vacuum of the main chamber is under $5\times10^{-10}$ Torr. As shown in **FIG. 4**, the relationship between the electron-beam current, emitted from the cathode, and the extraction voltage was measured with the magnetic-coil current of 20 A. The curve does not satisfy the Child's Law[66], a proportional relationship between the emission current and the 3/2 power of the extraction voltage, which may reflect the fact that the cathode was underheated. A 9.29 mA maximum electron-beam current was obtained with 1100 V extraction voltage. As displayed in **TABLE II**, when the electron-beam energy was set to 30 eV, 74.1% of the 1.12 mA emitted electron beam was collected by the electron collector, while most of the remaining electrons hit the anode. With a more than 80 eV electron-beam energy, the collection ratio was higher than 98.5%.

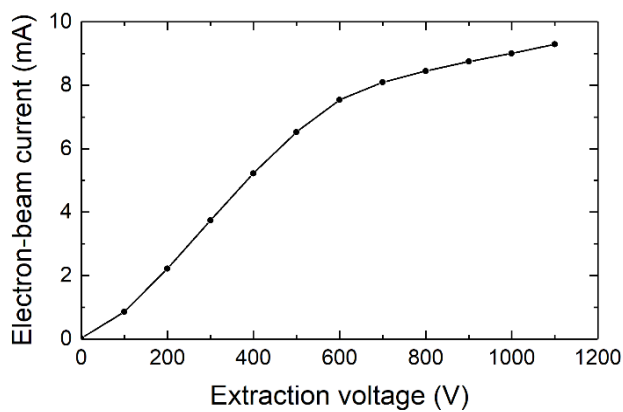

**FIG. 4**. A relationship between the electron-beam current, emitted from the cathode, and the extraction voltage (the voltage difference between the anode and the cathode). The maximum electron-beam current obtained with 1100 V extraction voltage was 9.29 mA.

**TABLE II.** The collection ratio of the electron beam at different electron-beam energy. It is noted that the electron-beam currents were not at their maximum values when the electron-beam energies were 604 eV and 1546 eV.

| Electron-beam energy (eV) | 30 | 60 | 80 | 150 | 604 | 1546 | 2500 |
|---|---|---|---|---|---|---|---|
| Emitted beam current (mA) | 1.12 | 2.09 | 3.15 | 8.42 | 6.84 | 6.25 | 9.29 |
| Collected beam current (mA) | 0.83 | 2.04 | 3.14 | 8.42 | 6.75 | 6.19 | 9.18 |
| Collection ratio (%) | 74.1 | 97.6 | 99.7 | 100 | 98.7 | 99.0 | 98.8 |

## B. HCIs extraction and charge-state distribution

In order to obtain a tungsten HCI beam, the hexacarbonyl tungsten [W(CO)$_6$] vapor was injected into the SW-EBIT, and the parameters set were similar to those used in the simulation in **FIG. 2**. On the pulse mode, a periodic pulse signal (pulse width: 100 μs) was amplified by a high voltage amplifier (Trek-2220) to raise the potential of DT2 within 10 μs rise time and to extract the tungsten HCIs produced and trapped in DT2. When the trap was opened, the potential of DT2 was 50 V higher than DT3 but lower than DT1, and thus the tungsten HCIs were extracted immediately and a pulsed HCIs bunch was obtained. When extracted, the ions of different charge states were separated by a commercial Wien filter[43] and collected by a Faraday cup, as illustrated in **FIG. 5**. The ions with different charge states are accelerated by the same potential difference between DT2 and the entrance of the Wien filter, which results in different speeds. The ions with a specific speed $v = E/B$ are balanced by the orthogonal uniform electric field $E$ and magnetic field $B$, which can pass the Wien filter directly, while the others are deflected. In our case, with the 0.5 T magnetic field and 2 mm aperture of the Wien filter, the charge-state distribution of tungsten ions was measured by scanning the voltage of the Wien filter. By triggering a picoammeter (Keithley 6485) which was connected to the Faraday cup, within the 0.4 ms integration time, the average current nearby the peak of the ion-beam current pulse (the pulse FWHM: ~1 ms) was measured.

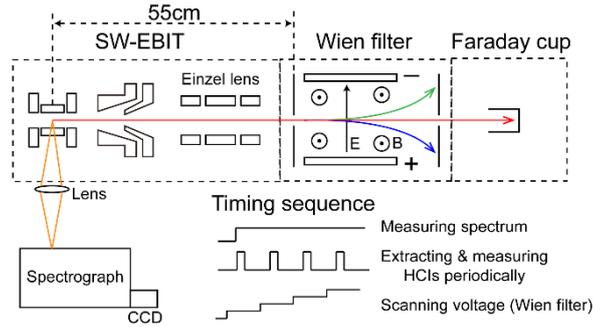

**FIG. 5**. Scheme for extracting and analyzing the HCI beam, and for measuring the spectra. The spectra are measured by an Andor Shamrock 303i spectrograph equipped with an Andor Newton 940 CCD from the slit of DT2.

When the uncalibrated electron-beam energy was set to be 100 eV, the charge-state distribution at different extraction period was measured, as seen in **FIG. 6**. When the extraction period was set to be 15 ms and 200 ms, the low charge-state tungsten ions were obtained. If the trap was kept opened, the peak of the CO$^+$ ion could be resolved. When the extraction period was 200 ms, a small peak of W$^{7+}$ ion arose in advance of the ionization energy of the ground state W$^{6+}$ (122.01 eV[67]) caused by the indirect ionization mechanism.[68]

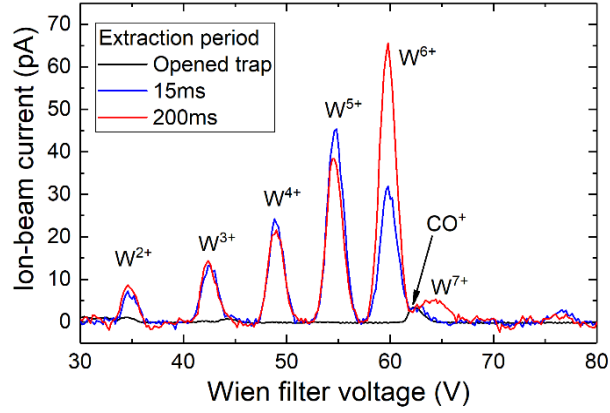

**FIG. 6**. Charge-state distributions of tungsten ions of 0 ms (the opened trap), 15 ms, and 200 ms extraction periods with 100 eV uncalibrated electron-beam energy. When the trap was kept opened, a DC ion-beam current was obtained and measured by the picoammeter with 100 ms integration time.

### C. Preliminary test for line identification

The $^{184}W^{14+}$ ion was proposed as one of the candidate ions for optical atomic clocks.[69,70] Here, the $W^{14+}$ ions of natural isotopic composition were produced and extracted. The wavelength of the clock transition of $^{184}W^{14+}$ is longer than 1000 nm.[69,70] We failed this line measurement from the SW-EBIT, due to the intense infrared radiation of the hot cathode. There is a disagreement about the identification of the line at 549.9 nm (in the air). In 2015, Zhao *et al.* identified this line to be an M1 transition between the ground state ($4f^{13}\,5s^2\,^2F$) fine structure levels of $W^{13+}$.[71] Later in the same year, Kobayashi *et al.* identified this line to be the transition of $W^{14+}$ based on a time-of-flight (TOF) measurement.[72] Therefore, it is interesting to measure this line in the SW-EBIT with the Wien filter.

The tungsten ions were produced by a 4.1 mA electron beam with the electron-beam energy of 279(4) eV and 300(4) eV, and were extracted at every 0.5 s during the spectrum measurements. The electron-beam energy was corrected for the space-charge effect described in Ref.73. As shown in **FIG. 7**, the spectra in the range of 530–580 nm and the charge-state distributions of the tungsten ions were measured. Comparing with **FIG. 7(b)**, the intensity of the 549.9 nm line behaves consistently with the number of extracted $W^{14+}$ ions, indicating that the line at 549.9 nm is most likely from $W^{14+}$. However, the intensities of the lines from $W^{12+}$ and $W^{13+}$ ions in **FIG. 7(a)** seem not very consistent with the numbers of the corresponding extracted ions, implying that more work needs to be done before reaching a definitive conclusion.

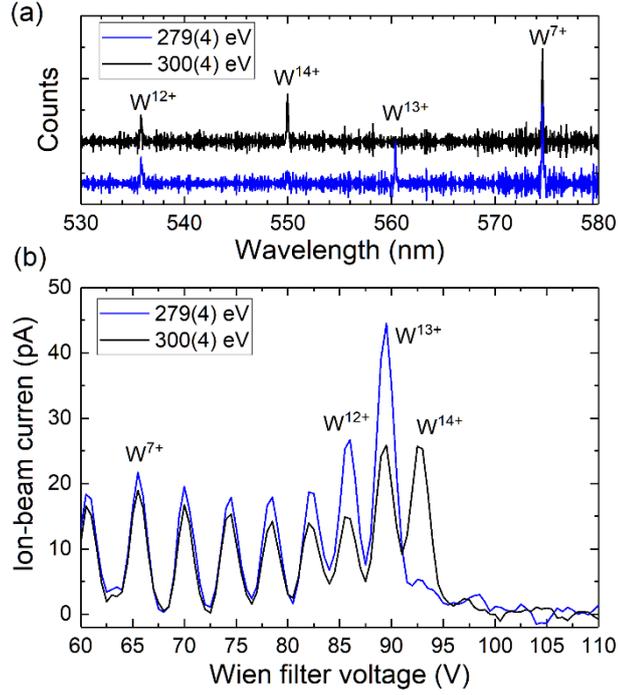

**FIG. 7**. (a) Spectra in the range of 530–580 nm (in the air) of tungsten ions of 279(4) eV and 300(4) eV electron-beam energies. The identifications refer to Ref. 72. (b) Charge-state distributions of tungsten ions of 279(4) eV and 300(4) eV electron-beam energies. The tungsten ions were extracted at every 0.5 s interval during spectrum measurements.

## IV. CONCLUSIONS

The low-energy, compact and high-temperature superconducting EBIT, the SW-EBIT, has been designed, built, and tested. The lower limit of the electron-beam energy is 30 eV with the 74.1% collection ratio of the 1.12 mA emitted electron beam. For electron-beam energy higher than 80 eV, the 98.5% collection ratio is guaranteed with the electron-beam current up to 9.29 mA. The low charge state tungsten ions have been produced, extracted, and separated by the Wien filter with the nominal electron-beam energy of 100 eV, which has demonstrated the potential of the SW-EBIT for performing research of low-charge-state HCIs. The charge-state distributions and spectra in the range of 530–580 nm for tungsten ions have been measured simultaneously with the electron energy of 279(4) eV and 300(4) eV, which has preliminarily confirmed the 549.9 nm line belonging to $W^{14+}$. These tests have shown that the SW-EBIT is capable of being a light source and an ion source of moderate-charge-state HCIs.

## ACKNOWLEDGMENTS

We appreciate Ming-Sheng Zhan for his suggestion and support of this project. This project was inspired by the pioneering work of Andrei Derevianko and coworkers,[62] and we thank his help and fruitful discussions. We were benefited greatly from a series of workshops on Precision Physics Based on Highly Charged Ions organized by Zong-Chao Yan, in which there were many fruitful discussions with and suggestions from the participants, including Andrei Derevianko, José R. Crespo López-Urrutia, B. K. Sahoo, Yan-Mei Yu, and Ji-Guang Li. This work is supported by the Strategic Priority Research Program of the Chinese Academy of Sciences (CAS) (Grant No. XDB21030300), the National Natural Science Foundation of China (Grant Nos. 11622434, 11704398), the Chinese National Fusion Project for ITER (Grant No. 2015GB117000), CAS Youth Innovation Promotion Association (Grant No. 2015274), and Hubei Province Science Fund for Distinguished Young Scholars (Grant No. 2017CFA040).

# REFERENCES

[1] M.A. Levine, R.E. Marrs, J.R. Henderson, D.A. Knapp, and M.B. Schneider, Phys. Scr. **T22**, 157 (1988).

[2] R.E. Marrs, M.A. Levine, D.A. Knapp, and J.R. Henderson, Phys. Rev. Lett. **60**, 1715 (1988).

[3] D.A. Knapp, R.E. Marrs, S.R. Elliott, E.W. Magee, and R. Zasadzinski, Nucl. Instruments Methods Phys. Res. Sect. A Accel. Spectrometers, Detect. Assoc. Equip. **334**, 305 (1993).

[4] R.E. Marrs, S.R. Elliott, and D.A. Knapp, Phys. Rev. Lett. **72**, 4082 (1994).

[5] J.D. Gillaspy, J. Phys. B At. Mol. Opt. Phys. **34**, R93 (2001).

[6] J.R. Henderson, P. Beiersdorfer, C.L. Bennett, S. Chantrenne, D.A. Knapp, R.E. Marrs, M.B. Schneider, K.L. Wong, G.A. Doschek, J.F. Seely, C.M. Brown, R.E. LaVilla, J. Dubau, and M.A. Levine, Phys. Rev. Lett. **65**, 705 (1990).

[7] N. Nakamura, H. Tobiyama, H. Nohara, A.P. Kavanagh, H. Watanabe, H.A. Sakaue, Y. Li, D. Kato, F.J. Currell, C. Yamada, and S. Ohtani, J. Phys. Conf. Ser. **58**, 267 (2007).

[8] D.A. Knapp, R.E. Marrs, M.A. Levine, C.L. Bennett, M.H. Chen, J.R. Henderson, M.B. Schneider, and J.H. Scofield, Phys. Rev. Lett. **62**, 2104 (1989).

[9] D.A. Knapp, Zeitschrift Für Phys. D Atoms, Mol. Clust. **21**, S143 (1991).

[10] D.A. Knapp, R.E. Marrs, M.B. Schneider, M.H. Chen, M.A. Levine, and P. Lee, Phys. Rev. A **47**, 2039 (1993).

[11] D.A. Knapp, P. Beiersdorfer, M.H. Chen, J.H. Scofield, and D. Schneider, Phys. Rev. Lett. **74**, 54 (1995).

[12] P. Beiersdorfer, E. Träbert, H. Chen, M.-H. Chen, M.J. May, and A.L. Osterheld, Phys. Rev. A **67**, 052103 (2003).

[13] A. Gumberidze, T. Stöhlker, D. Banaś, K. Beckert, P. Beller, H.F. Beyer, F. Bosch, S. Hagmann, C. Kozhuharov, D. Liesen, F. Nolden, X. Ma, P.H. Mokler, M. Steck, D. Sierpowski, and S. Tashenov, Phys. Rev. Lett. **94**, 223001 (2005).

[14] P. Beiersdorfer, J. Phys. B At. Mol. Opt. Phys. **43**, 074032 (2010).

[15] J.D. Silver, A.J. Varney, H.S. Margolis, P.E.G.G. Baird, I.P. Grant, P.D. Groves, W.A. Hallett, A.T. Handford, P.J. Hirst, A.R. Holmes, D.J.H.H. Howie, R.A. Hunt, K.A. Nobbs, M. Roberts, W. Studholme, J.S. Wark, M.T. Williams, M.A. Levine, D.D. Dietrich, W.G. Graham, I.D. Williams, R. O'Neil, and S.J. Rose, Rev. Sci. Instrum. **65**, 1072 (1994).

[16] S. Böhm, A. Enulescu, T. Fritioff, I. Orban, S. Tashenov, and R. Schuch, J. Phys. Conf. Ser. **58**, 303 (2007).

[17] Y. Fu, K. Yao, B. Wei, D. Lu, R. Hutton, and Y. Zou, J. Instrum. **5**, C08011 (2010).

[18] C.A. Morgan, F.G. Serpa, E. Takács, E.S. Meyer, J.D. Gillaspy, J. Sugar, J.R. Roberts, C.M. Brown, and U. Feldman, Phys. Rev. Lett. **74**, 1716 (1995).

[19] F.J. Currell, J. Asada, K. Ishii, A. Minoh, K. Motohashi, N. Nakamura, K. Nishizawa, S. Ohtani, K. Okazaki, M. Sakurai, H. Shiraishi, S. Tsurubuchi, and H. Watanabe, J. Phys. Soc. Japan **65**, 3186 (1996).

[20] C. Biedermann, A. Förster, G. Fußmann, and R. Radtke, Phys. Scr. **T73**, 360 (1997).

[21] N. Nakamura, J. Asada, F.J. Currell, T. Fukami, T. Hirayama, K. Motohashi, T. Nagata, E. Nojikawa, S. Ohtani, K. Okazaki, M. Sakurai, H. Shiraishi, S. Tsurubuchi, and H. Watanabe, Phys. Scr. **T73**, 362 (1997).

[22] J.R. Crespo López-Urrutia, A. Dorn, R. Moshammer, J. Ullrich, J.R.C. L?pez-Urrutia, A. Dorn, R. Moshammer, and J. Ullrich, Phys. Scr. **T80**, 502 (1999).

[23] V.P. Ovsyannikov and G. Zschornack, Rev. Sci. Instrum. **70**, 2646 (1999).

[24] H. Watanabe and F. Currell, J. Phys. Conf. Ser. **2**, 182 (2004).

[25] M. He, Y. Liu, Y. Yang, S. Wu, W. Chen, W. Hu, P. Guo, D. Lu, Y. Fu, M. Huang, X. Zhang, R. Hutton, L. Liljeby, and Y. Zou, J. Phys. Conf. Ser. **58**, 419 (2007).

[26] P. Beiersdorfer, A.L. Osterheld, M.H. Chen, J.R. Henderson, D.A. Knapp, M.A. Levine, R.E. Marrs, K.J. Reed, M.B. Schneider, and D.A. Vogel, Phys. Rev. Lett. **65**, 1995 (1990).

[27] J.T. Jefferies, F.Q. Orrall, and J.B. Zirker, Sol. Phys. **16**, 103 (1971).

[28] R. Radtke, C. Biedermann, P. Mandelbaum, and J.L. Schwob, J. Phys. Conf. Ser. **58**, 113 (2007).

[29] G.Y. Liang, T.M. Baumann, J.R. Crespo López-Urrutia, S.W. Epp, H. Tawara, A. Gonchar, P.H. Mokler, G. Zhao, and J. Ullrich, Astrophys. J. **696**, 2275 (2009).

[30] P. Beiersdorfer, Annu. Rev. Astron. Astrophys. **41**, 343 (2003).

[31] H. Bekker, C. Hensel, A. Daniel, A. Windberger, T. Pfeifer, and J.R. Crespo López-Urrutia, Phys. Rev. A **98**, 062514 (2018).

[32] N. Nakamura, H. Kikuchi, H.A. Sakaue, and T. Watanabe, Rev. Sci. Instrum. **79**, 063104 (2008).

[33] N. Yamamoto, H.A. Sakaue, D. Kato, I. Murakami, T. Kato, N. Nakamura, E. Watanabe, H. Nishimura, and T. Watanabe, J. Phys. Conf. Ser. **163**, 012023 (2009).

[34] S.W. Epp, J.R. Crespo López-Urrutia, M.C. Simon, T. Baumann, G. Brenner, R. Ginzel, N. Guerassimova, V. Mäckel, P.H. Mokler, B.L. Schmitt, H. Tawara, and J. Ullrich, J. Phys. B At. Mol. Opt. Phys. **43**, 194008 (2010).

[35] J. Xiao, R. Zhao, X. Jin, B. Tu, Y. Yang, D. Lu, R. Hutton, and Y. Zou, in Proc. IPAC (JACoW, 2013), p. 434; available at http://accelconf.web.cern.ch/AccelConf/IPAC2013/papers/mopfi066.pdf .

[36] J. Xiao, Z. Fei, Y. Yang, X. Jin, D. Lu, Y. Shen, L. Liljeby, R. Hutton, and Y. Zou, Rev. Sci. Instrum. **83**, 013303 (2012).

[37] J.K. Lepson and P. Beiersdorfer, Phys. Scr. **T120**, 62 (2005).

[38] R. Radtke, C. Biedermann, J.L. Schwob, P. Mandelbaum, and R. Doron, Phys. Rev. A **64**, 012720 (2001).

[39] E. Träbert, P. Beiersdorfer, S.B. Utter, and J.R. Crespo López-Urrutia, Phys. Scr. **58**, 599 (1998).

[40] J.K. Lepson, P. Beiersdorfer, G. V. Brown, D.A. Liedahl, S.B. Utter, N.S. Brickhouse, A.K. Dupree, J.S. Kaastra, R. Mewe, and S.M. Kahn, Astrophys. J. **578**, 648 (2002).

[41] D. Schneider, M.W. Clark, B.M. Penetrante, J. McDonald, D. DeWitt, and J.N. Bardsley, Phys. Rev. A **44**, 3119 (1991).

[42] G. Zschornack, M. Kreller, V.P. Ovsyannikov, F. Grossman, U. Kentsch, M. Schmidt, F. Ullmann, and R. Heller, Rev. Sci. Instrum. **79**, 02A703 (2008).

[43] M. Schmidt, H. Peng, G. Zschornack, and S. Sykora, Rev. Sci. Instrum. **80**, 063301 (2009).

[44] U. Kentsch, G. Zschornack, A. Schwan, and F. Ullmann, Rev. Sci. Instrum. **81**, 02A507 (2010).

[45] G. Zschornack, J. König, M. Schmidt, and A. Thorn, Rev. Sci. Instrum. **85**, 02B703 (2014).

[46] S.F. Hoogerheide and J.N. Tan, J. Phys. Conf. Ser. **583**, 012044 (2015).

[47] V.P. Ovsyannikov and A.V. Nefiodov, Nucl. Instruments Methods Phys. Res. Sect. B Beam Interact.

with Mater. Atoms **370**, 32 (2016).

[48] P. Micke, S. Kühn, L. Buchauer, J.R. Harries, T.M. Bücking, K. Blaum, A. Cieluch, A. Egl, D. Hollain, S. Kraemer, T. Pfeifer, P.O. Schmidt, R.X. Schüssler, C. Schweiger, T. Stöhlker, S. Sturm, R.N. Wolf, S. Bernitt, and J.R. Crespo López-Urrutia, Rev. Sci. Instrum. **89**, 063109 (2018).

[49] M.A. Blessenohl, S. Dobrodey, C. Warnecke, M.K. Rosner, L. Graham, S. Paul, T.M. Baumann, Z. Hockenbery, R. Hubele, T. Pfeifer, F. Ames, J. Dilling, and J.R. Crespo López-Urrutia, Rev. Sci. Instrum. **89**, 052401 (2018).

[50] G. Kraft, in *Phys. Mult. Highly Charg. Ions Vol. 1. Sources, Appl. Fundam. Process.*, edited by F.J. Currell (Springer, Dordrecht, The Netherlands, 2003), pp. 149–196.

[51] L. Gruber, J.P. Holder, B.R. Beck, J. Steiger, J.W. McDonald, J. Glassman, H. DeWitt, D.A. Church, and D. Schneider, *Hyperfine Interact.* **127**, 215 (2000).

[52] L. Gruber, J.P. Holder, J. Steiger, B.R. Beck, H.E. DeWitt, J. Glassman, J.W. McDonald, D.A. Church, and D. Schneider, Phys. Rev. Lett. **86**, 636 (2001).

[53] A.D. Ludlow, M.M. Boyd, J. Ye, E. Peik, and P.O. Schmidt, Rev. Mod. Phys. **87**, 637 (2015).

[54] J.C. Berengut, V.A. Dzuba, and V.V. Flambaum, Phys. Rev. Lett. **105**, 120801 (2010).

[55] A. Derevianko, V.A. Dzuba, and V.V. Flambaum, Phys. Rev. Lett. **109**, 180801 (2012).

[56] M.S. Safronova, V.A. Dzuba, V.V. Flambaum, U.I. Safronova, S.G. Porsev, and M.G. Kozlov, Phys. Rev. Lett. **113**, 030801 (2014).

[57] L.S. Brown and G. Gabrielse, Rev. Mod. Phys. **58**, 233 (1986).

[58] M. Schwarz, O.O. Versolato, A. Windberger, F.R. Brunner, T. Ballance, S.N. Eberle, J. Ullrich, P.O. Schmidt, A.K. Hansen, A.D. Gingell, M. Drewsen, J.R. Crespo López-Urrutia, and J.R.C. López-Urrutia, Rev. Sci. Instrum. **83**, 083115 (2012).

[59] L. Schmöger, M. Schwarz, T.M. Baumann, O.O. Versolato, B. Piest, T. Pfeifer, J. Ullrich, P.O. Schmidt, and J.R. Crespo López-Urrutia, Rev. Sci. Instrum. **86**, 103111 (2015).

[60] L. Schmoger, O.O. Versolato, M. Schwarz, M. Kohnen, A. Windberger, B. Piest, S. Feuchtenbeiner, J. Pedregosa-Gutierrez, T. Leopold, P. Micke, A.K. Hansen, T.M. Baumann, M. Drewsen, J. Ullrich, P.O. Schmidt, and J.R. Crespo López-Urrutia, Science **347**, 1233 (2015).

[61] P.O. Schmidt, Science **309**, 749 (2005).

[62] V.I. Yudin, A.V. Taichenachev, and A. Derevianko, Phys. Rev. Lett. **113**, 233003 (2014).

[63] Y.M. Yu and B.K. Sahoo, Phys. Rev. A **94**, 62502 (2016).

[64] Y.M. Yu and B.K. Sahoo, Phys. Rev. A **97**, 041403(R) (2018)..

[65] COMSOL Multiphysics Reference Manual 5.3a, COMSOL, Inc., 2017.

[66] C.D. Child, Phys. Rev. (Series I) **32**, 492 (1911).

[67] A. Kramida, Yu. Ralchenko, J. Reader, and NIST ASD Team (2018). *NIST Atomic Spectra Database* (ver. 5.6.1), [Online]. Available: https://physics.nist.gov/asd [2019, May 30]. National Institute of Standards and Technology, Gaithersburg, MD. DOI: https://doi.org/10.18434/T4W30F

[68] Q. Lu, J. He, H. Tian, M. Li, Y. Yang, K. Yao, C. Chen, J. Xiao, J.G. Li, B. Tu, and Y. Zou, Phys. Rev. A **99**, 042510 (2019).

[69] V.A. Dzuba, A. Derevianko, and V.V. Flambaum, Phys. Rev. A **86**, 054501 (2012).

[70] A. Windberger, J.R. Crespo López-Urrutia, H. Bekker, N.S. Oreshkina, J.C. Berengut, V. Bock, A. Borschevsky, V.A. Dzuba, E. Eliav, Z. Harman, U. Kaldor, S. Kaul, U.I. Safronova, V.V. Flambaum, C.H. Keitel, P.O. Schmidt, J. Ullrich, and O.O. Versolato, Phys. Rev. Lett. **114**, 150801 (2015).

[71] Z.Z. Zhao, M.L. Qiu, R.F. Zhao, W.X. Li, X.L. Guo, J. Xiao, C.Y. Chen, Y. Zou, and R. Hutton, J.

Phys. B At. Mol. Opt. Phys. **48**, 115004 (2015).

[72] Y. Kobayashi, K. Kubota, K. Omote, A. Komatsu, J. Sakoda, M. Minoshima, D. Kato, J. Li, H.A. Sakaue, I. Murakami, and N. Nakamura, Phys. Rev. A **92**, 022510 (2015).

[73] B. Tu, Q.F. Lu, T. Cheng, M.C. Li, Y. Yang, K. Yao, Y. Shen, D. Lu, J. Xiao, R. Hutton, and Y. Zou, Phys. Plasmas **24**, 103507 (2017).